\documentclass{emulateapj}
\shorttitle{EUV Late Phase of a Flare}
\shortauthors{Dai et al.}
\slugcomment{Accepted for publication in the Astrophysical Journal Letters}
\begin{document}

\title{Production of the Extreme-Ultraviolet Late Phase of an X Class Flare in a Three-Stage 
Magnetic Reconnection Process}

\author{Y.~Dai\altaffilmark{1,2}, M.~D.~Ding\altaffilmark{1,2}, and Y.~Guo\altaffilmark{1,2}}

\affil{$^{1}$School of Astronomy and Space Science, Nanjing University, Nanjing 210093, China; \url{ydai@nju.edu.cn}\\
$^{2}$Key Laboratory of Modern Astronomy and Astrophysics (Nanjing University), 
Ministry of Education, Nanjing 210093, China}


\begin{abstract}
We report observations of an X class flare on 2011 September 6 by the instruments onboard
the \emph{Solar Dynamics Observatory} (\emph{SDO}).  The flare occurs in a complex active region with multiple polarities. 
The Extreme-Ultraviolet (EUV) Variability Experiment (EVE) observations in the warm coronal emission reveal three enhancements, 
of which the third one corresponds to an EUV  late phase. The three enhancements have a one-to-one correspondence to the 
three stages in flare evolution identified by the spatially-resolved Atmospheric Imaging Assembly (AIA) observations, which
are characterized by a flux rope eruption, a moderate filament ejection, and the appearance of EUV late phase 
loops, respectively. The EUV late phase loops are spatially and morphologically distinct from the main flare loops. 
Multi-channel analysis suggests the presence of a continuous but fragmented energy injection during the EUV late phase resulting
in the warm corona nature of the late phase loops. Based on these observational facts, We propose a three-stage magnetic 
reconnection scenario to explain the flare evolution. Reconnections in different stages involve different magnetic fields but 
show a casual relationship between them. The EUV late phase loops are mainly produced by the least energetic magnetic
reconnection in the last stage.
\end{abstract}


\section{INTRODUCTION}

It is widely accepted that flares are a result of the rapid release of magnetic energy stored in the solar corona. Among the radiative
output from a flare, emission from the Extreme-Ultraviolet (EUV) spectral range covers a substantial fraction, thus serving as 
an important tool to diagnose the dynamics and evolution of the flare. Observations of the Sun in EUV have been carried out
for more than half a century \citep[e.g.,][]{Firedman63}. However, previous observations often suffered from the limited 
spectral range. The situation has been greatly improved since the recently launched
\emph{Solar Dynamics Observatory} \citep[\emph{SDO};][]{Pesnell12} mission provides both spectroscopic observations with
full EUV spectral coverage and simultaneous imaging observations in multiple EUV bandpasses. 

One of the intriguing phenomena discovered by \emph{SDO} is an ``EUV late phase" in some flares \citep{Woods11}, which is
seen as  a second peak in the warm coronal ($\sim$3 MK) emissions  (such as Fe~{\scriptsize XVI}) several minutes to a few hours
after the \emph{GOES} soft X-ray (SXR) peak. There are, however, no significant enhancements of the SXR or hot coronal 
($\sim$10 MK) emissions in the EUV late phase, and spatially-resolved observations show that the secondary warm coronal 
emission comes from a set of longer loops rather than the original flaring loops. Up to now, there are only a few reports of the EUV
late phase in literatures, and the origin of the EUV late phase is still not fully understood. Some authors thought that the EUV late 
phase is due to a second energy injection late in the flare \citep[e.g.,][]{Woods11,Hock12b}, while others proposed that it
is mainly a cooling-effect extending from the initial main flare heating \citep[e.g.,][]{Liu13}. To clarify this question,
in this Letter we present \emph{SDO} observations of an X class flare on 2011 September 6, which exhibits an extended EUV
late phase. Based on the observational facts found in this flare, we propose a specific scenario, three-stage magnetic reconnection,
to explain the production of the EUV late phase.

\section{OBSERVATIONS AND DATA ANALYSIS} 
We use data from the three instruments onboard \emph{SDO}\@.  The EUV Variability Experiment \citep[EVE;][]{Woods12} 
measures full-disk integrated EUV irradiance from 0.1 to 105 nm with 0.1 nm spectral resolution  and 10 s temporal cadence.
The EVE data used in this study are primarily from the MEGS-A channel \citep{Hock12a},
which covers the 7--37 nm wavelength range with a nearly 100\% duty cycle. The Atmospheric Imaging Assembly 
\citep[AIA;][]{Lemen12} provides simultaneous full-disk images of the transition region and corona  in 10 channels with 
1$\farcs$5 spatial resolution and 12 s temporal resolution. In addition, magnetograms from the Helioseismic and Magnetic Imager 
\citep[HMI;][]{Scherrer12} are chosen to check the magnetic topology of the flare. 

\begin{figure}
\epsscale{1}
\plotone{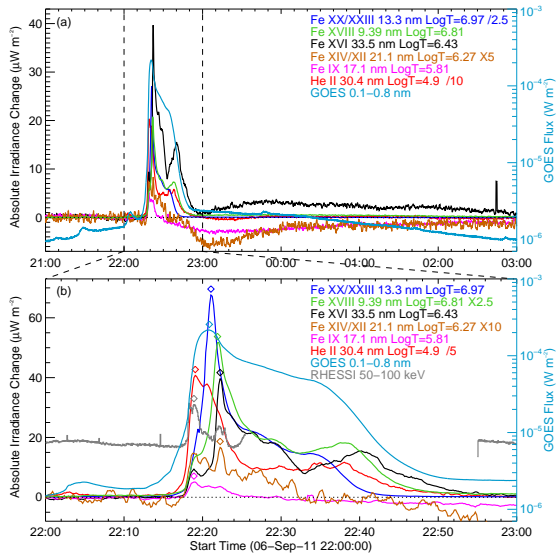}
\caption{Time profiles of the background-subtracted irradiance in six \emph{SDO}/EVE spectral lines and  \emph{GOES}
0.1--0.8 nm flux for the  2011 September 6  X2.1 flare. Panel (a) shows the whole evolution of the flare,  during which the 
warm  coronal Fe {\scriptsize XVI} 33.5 nm emission exhibits  an extended late phase lasting until the first few hours 
of 2011 September 7. Panel (b) gives an enlarged view of the flare evolution between 22:00 UT and 23:00 UT,  with the 
\emph{RHESSI}  50--100 keV count rate over-plotted. The color-coded  diamonds denote the peak time
for the corresponding emission. }
\end{figure}

The flare under study occurred in NOAA active region (AR) 11283 on 2011 September 6, positioned close to the 
disk center. 
It is an X2.1 class flare that started at 22:12 UT, peaked 
around 22:21 UT, and ended at 22:24 UT, as revealed by the \emph{GOES} 0.1--0.8 nm light curve in Figure 1. 
 
To study the thermal evolution of the flare, in Figure 1 we also plot the  time profiles of the background-subtracted
irradiance in six EVE spectral lines, which cover temperatures from $\log T\sim4.9$ to $\log T\sim7$. The flare 
exhibits the all four characteristics pointed out in \citet{Woods11}. First, the cold chromospheric He~{\scriptsize II} 30.4 nm 
emission showed an impulsive enhancement and peaked around 22:19 UT, almost coincident with the \emph{RHESSI}
50--100 keV hard X-ray (HXR) emission peak, as well as the spikes in the emissions from Fe~{\scriptsize IX} to 
Fe~{\scriptsize XVI}. This implies that  the chromosphere, as responding to impact of the non-thermal electrons accelerated 
during the flare's impulsive phase, was instantly heated to $\log T\sim6.4$  \citep{Milligan12,Chamberlin12}. Second, the hot coronal 
Fe~{\scriptsize XX/XXIII} 13.3 nm emission closely resembled the \emph{GOES} SXR time series and peaked around 22:21 UT, 
which is believed to be  a result of the chromospheric evaporation caused by the initial heating. The cooler emissions then peaked
sequentially with decreasing temperatures within a short period of  70 s, revealing a fast cooling rate over $1\times10^5$ K s$^{-1}$. 
Third, the cool coronal Fe~{\scriptsize IX} 17.1 nm and moderately warm coronal Fe~{\scriptsize XIV/XII} 21.1 nm emissions
decreased after the peak of the main phase and turned into a coronal dimming.  A coronal mass ejection (CME) was observed being
associated with the flare, and the coronal dimming is most likely to reflect the mass  drainage during the CME launch 
\citep[e.g.,][]{Aschwanden09}. Fourth,  as the dimming in cool coronal emissions developed and other  emissions returned to the 
pre-flare level, the warm coronal  Fe~{\scriptsize XVI} 33.5 nm emission showed  another small enhancement that started from 
$\sim$23:00 UT and lasted until the first few hours of 2011 September 7. It  should correspond to the EUV late phase as 
defined in \citet{Woods11}, which will be further validated by the  aftermentioned AIA imaging observations. Note that in this late 
phase there was also likely a very weak increase in the moderately hot coronal Fe~{\scriptsize XVIII} 9.39 nm emission. 

Besides these four features, we find that additional moderate enhancements in the EVE
30.4, 33.5, and 9.39 nm emissions were seen to start from $\sim$22:33 UT and peak around 22:40 UT\@. The EVE
13.3 nm emission and the \emph{GOES} SXR  flux also showed a hump during this period. Therefore, compared to the typical 
EUV late phase as demonstrated in \citet{Woods11},  which is directly preceded by the flare's main phase, in this flare there is 
another ``intermediate phase" characterized by the moderate peak between the main phase and the late phase.  

\begin{figure*}
\epsscale{1}
\plotone{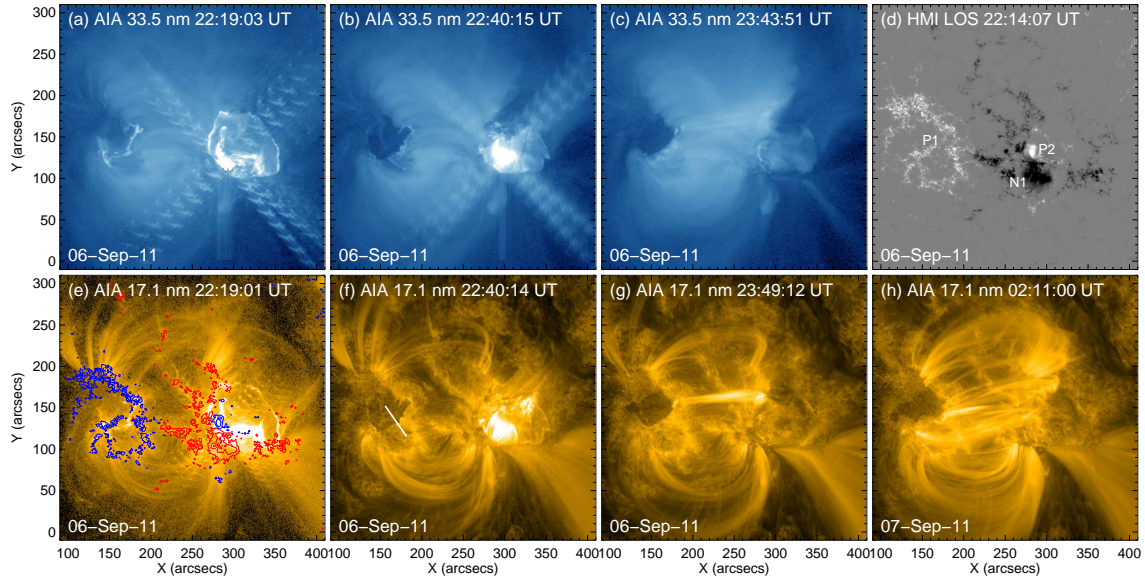}
\caption{Snapshots of  the flare evolution taken by \emph{SDO}/AIA at 33.5 (a--c) and 17.1 nm (e--h), as well as a 
light-of-sight (LOS) magnetogram of AR 11283 taken at 22:14 UT by \emph{SDO}/HMI (d). Three main polarities are
identified and labeled as P1, N1, and P2, whose contours are also overlaid in panel (e).  A slice is placed in panel (f), 
along which the evolution will be traced in Figure 3.}
\end{figure*}

Spatially-resolved AIA observations of the flare evolution in six coronal channels are presented in the online animation. Figure 2 
shows some snapshots of the animation in the AIA 33.5 and 17.1 nm channels, respectively, as well as a  light-of-sight (LOS)
magnetogram of AR 11283 taken at 22:14 UT by HMI\@. Three main polarities are identified and labeled as P1, N1, and P2,
among which P2 is a parasitic positive polarity embedded in the negative polarity N1.  The flare first occurred along the 
southeastern part of the polarity inversion line (PIL) between N1 and P2, behaving as a sigmoid-to-flux-rope eruption pattern
\citep[cf,][]{Liu10}, which is best seen in AIA 13.1 and 9.4 nm. Note that prior to the event a semi-circular filament 
was observed to lie along the PIL, the southern part of which has been successfully modeled as a flux rope (FR) by
\citet{Feng13} and \citet{Jiang13} by using different nonlinear force-free field (NLFFF) extrapolation methods. From $\sim$22:18 UT
the FR started to rise rapidly, with the flare ribbons becoming more elongated along the PIL and the outer one eventually turning into 
a circular ribbon (Figures 2(a) and (e)).  Following the brightening of the main flare ribbons, a remote flare ribbon appeared in P1,
over a distance of 80$^{\prime\prime}$ east of the main flare region. The situation is very similar to those studied in 
\citet{Masson09} and \citet{Wang12}, indicating the presence of a 3D null-point magnetic topology, which is also expected from 
the LOS magnetogram. Afterwards, the region on the eastern side of the remote ribbon quickly turned into a
coronal dimming that persisted during the whole evolution of the flare. 

This first stage eruption was followed by a second stage eruption, which was seen as the continuous ejection of cold material from
the northern part of the filament starting from $\sim$22:33 UT (Figures 2(b) and (f)). This moderate ejection produced
new post-flare loops mainly on its southern side rather than centered on the filament itself. In AIA 13.1 and 9.4 nm (see
online animation), there appeared a second set of longer but less prominent loops connecting the eastern side of
the filament (in N1) and the abovementioned remote flare ribbon (in P1), which were not visible in the cooler coronal 
channels at that moment. 

The second stage eruption lasted about 30 minutes. As the post-flare loops in the main flare region cooled down, longer
post-flare loops connecting N1 and P1 became more and more dominant in particular in the warm and cool coronal channels. 
The appearance of these loops coincided
with the enhancement in the EVE Fe {\scriptsize XVI} 33.5 nm emission starting from $\sim$23:00 UT, and the loops were
spatially distinct from the original post-flare loops, further supporting the identification of an EUV late phase for this third stage 
evolution. These late phase loops were first localized at the same location of the second set of loops previously only seen in the 
hotter coronal channels during the second stage (Figures 2(c) and (g)), later became more spatially spread (Figure 2(h)).  
As noted in \citet{Woods11}, the late phase loops are very diffuse in morphology in the warm coronal AIA 33.5 nm channel, but 
quite distinct in the cool coronal channels such as AIA 17.1 nm.  

\begin{figure}
\epsscale{1}
\plotone{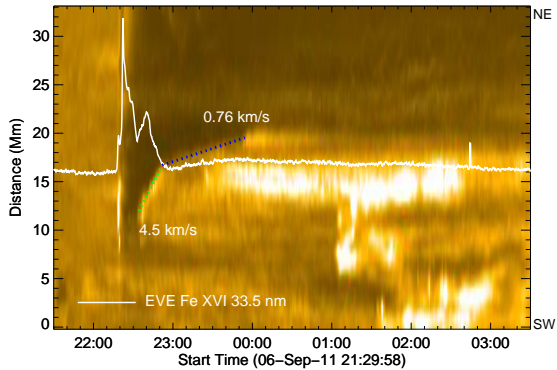}
\caption{Base ratio time-distance diagram of the AIA 17.1 nm images along the slice in Figure 2(f), showing the 
eastward (dimming-ward) expansion of the remote flare ribbon. The transition of the expansion velocity from 4.5 km s$^{-1}$ 
(green dotted line) to 0.76 km s$^{-1}$ (blue dotted line) coincides with the over-plotted EVE 33.5 nm emission variation
(white solid line).}
\end{figure}

It is worth noting that the remote flare ribbon brightened again and showed a long-lasting eastward (dimming-ward) expansion 
starting from the second stage eruption when long loops were found to anchor on it. This is an indicator of progressive magnetic
reconnection according to the standard flare model. In Figure 2(f) we place a slice roughly perpendicular to the remote ribbon
to trace its movement in AIA 17.1 nm. Because of the low brightness of the remote ribbon, we simply use linear fit to track 
the ribbon expansion. As revealed in Figure 3, the expansion velocity of the remote ribbon turned from 4.5 
km s$^{-1}$ to 0.76 km s$^{-1}$, coinciding with the transition from the second stage to the third stage, and also consistent with
the evolution of the EVE 
 33.5 nm emission. The expansion velocities are considerably lower than those 
in some typical two-ribbon flares, which are several tens of km s$^{-1}$ \citep[e.g.,][]{Asai04,Miklenic07}, implying less and less 
energetic magnetic reconnections during the second and third stages with much lower reconnection rates than the main flare 
reconnection.
 
\begin{figure*}
\epsscale{1}
\plotone{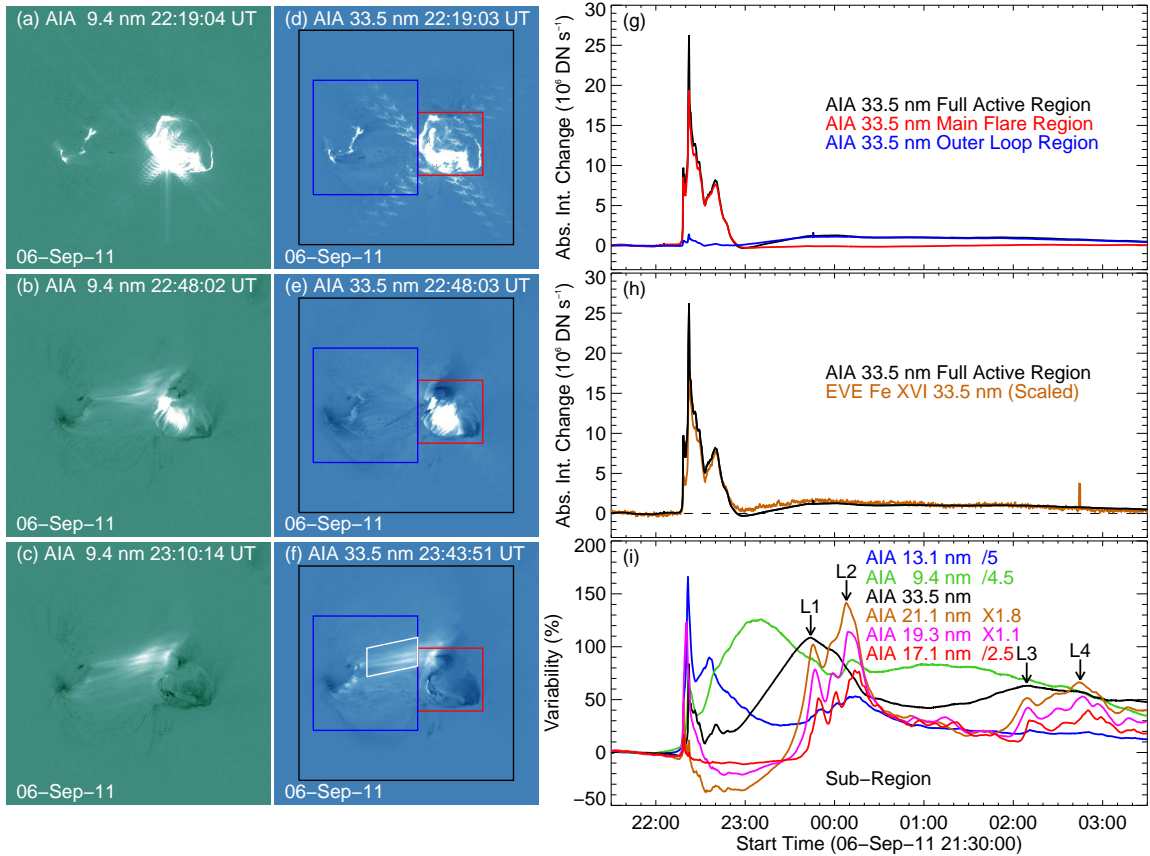}
\caption{Left and middle: AIA base difference images of the flare at 9.4 and 33.5 nm, respectively. In the right, from top to bottom: 
AIA 33.5 nm light curves from regions indicated by the color boxes in the AIA 33.5 nm images, comparison between the AIA 33.5 
nm full AR light curve and the EVE 33.5 nm profile, and intensity variability in six AIA coronal channels for 
the sub-region of the late phase loops defined by the white parallelogram in panel (f). Here the variability is defined as the relative
change from the background. Four episodes in the late phase are picked up by the arrows in panel (i) and labeled as L1--L4.}
\end{figure*}

The AIA base difference images in Figure 4 highlight the flare's three-stage evolution more clearly. To determine the 
contribution of different parts of the AR to the flare emission, we identify three regions, which are
indicated by the color boxes in the AIA 33.5 nm images, and calculate the light curve by summing the count rate
over all pixels in each region. The red box represents the main flare region, the blue one outlines the outer loop region, which is
believed to account for the EUV late phase,  and the black one surrounds the full AR\@. We plot the background-subtracted AIA 
light curves in 33.5 nm from the three regions in  Figure 4(g), and compare the AIA 33.5 nm full AR light curve with the EVE 
33.5 nm profile in Figure 4(h).  First, the general similarity between the two profiles in Figure 4(h) suggests that
the change in the EVE full-disk integrated irradiance comes mainly from the AR\@. Actually, there was no other major 
activity on the visible disk during the period of this event. Second, as expected, emission from the main flare region dominates in 
the main phase and the intermediate phase, while the outer loop region is responsible for the late phase. Note that in the main and 
intermediate phases there were also some intensity increases  from the outer loop region, which we attribute to
the brightening  of the remote flare ribbon and  CCD-bleed interference stripes extending from the intensively flaring site.
 
To study the thermal evolution the EUV late phase loops, we select a sub-region defined by the white parallelogram in Figure 4(f), 
which includes the most prominent late phase loops seen in AIA 33.5 nm, and calculate the intensity variability in six AIA coronal
channels for this region. As revealed in Figure 4(i), the EUV late phase can be decomposed into many episodes; each episode
is characterized by a cooling process, with the emissions peaking sequentially with decreasing temperatures.  Furthermore, 
combined with the imaging observations, it is found that these multiple peaks appear to be related to 
different, but adjacent, late phase loops cooling into the corresponding emission temperature ranges at different times, in
particular in the cool coronal channels. This indicates the presence of a continuous, but fragmented both temporally and spatially, 
energy injection during the EUV late phase, challenging the conclusion
of \citet{Liu13} that the appearance of the late phase loops is mainly a cooling-effect rather than the result of a later energy injection. 
We pick up four most prominent episodes labeled as L1--L4, and list the  time of peak emission seen in six AIA channels for each 
episode in Table 1. It is seen that in L2 the AIA 9.4 nm peak follows the AIA 19.3 nm peak, and all detectable peaks in AIA 13.1 nm 
are delayed from the corresponding AIA 17.1 nm peaks. This behavior can be expected from the AIA temperature response
functions (TRFs) calculated using the latest CHIANTI atomic database \citep{Landi13}, which also show a second cool coronal peak 
at $\log T\sim6.0$  for AIA 9.4 nm and $\log T\sim5.8$ for AIA 13.1 nm, and suggests the warm corona nature of the late phase 
loops. Between L2 and L3 the variabilities in AIA 9.4 and 33.5 nm showed opposite  patterns, further suggesting that during
this period the AIA 9.4 nm channel is mainly sensitive to cool late phase loops that cool from warm coronal temperatures.  
L1 is a little bit special, in which the appearance of  the late phase loops in AIA 33.5 nm (Figure 4(f)) is a combined effect of
the cooling from the hotter AIA 9.4 nm loops (Figure 4(c)) and the late phase energy injection. The cooling rates in these four 
episodes are 0.3--1.2 $\times10^4$ K s$^{-1}$, orders of magnitude lower than that in the main phase derived from the EVE
observations.

\begin{deluxetable*}{llccccc}
\tablecolumns{8}
\tablewidth{0pc}
\tablecaption{Time of the Peak Emission in AIA for the Selected Episodes in the Late Phase}
\tablehead{
\multicolumn{1}{c}{Channel} & \colhead{Ion} & \colhead{$\log T$} & \multicolumn{4}{c}{Peak Time (UT)}\\
\cline{4-7}\\
\multicolumn{1}{c}{(nm)} & \colhead{} & \colhead{}  & \colhead{L1 (6-Sep)} & 
\colhead{L2 (7-Sep)} & \colhead{L3 (7-Sep)} & \colhead{L4 (7-Sep)}
}
\startdata
9.4 (hot) & Fe {\scriptsize XVIII} & 6.8  & 23:10:14 & \nodata & \nodata & \nodata\\
33.5 & Fe {\scriptsize XVI} & 6.4  & 23:43:51 & \nodata & 02:09:27 & \nodata\\
21.1 & Fe {\scriptsize XIV} & 6.3   & 23:45:48 & 00:07:48 & 02:09:24 & 02:45:48\\
19.3 & Fe {\scriptsize XII} & 6.1  &23:46:55 & 00:09:19 & 02:09:55 & 02:48:31\\
9.4 (cool) & Fe {\scriptsize X} & 6.0 & \nodata & 00:12:02 & \nodata & \nodata\\
17.1 & Fe {\scriptsize IX} & 5.9  & 23:49:12  & 00:13:36 & 02:11:00 & 02:50:24\\
13.1 (cool)  & Fe {\scriptsize VIII} & 5.8  & 23:51:21 & 00:13:45 & 02:11:21 & \nodata 
\enddata
\end{deluxetable*}

\section{DISCUSSION AND CONCLUSION}
The 2011 September 6 X2.1 flare  conforms to the criteria for an EUV late phase flare defined in \citet{Woods11}. To our knowledge, 
this may be the first report of the EUV late phase of a flare above the X class. The ratio of the late phase peak to the main phase
peak is only 9.4\%, significantly lower than those in \citet{Woods11}. We attribute this low ratio to the intense
energy injection during the impulsive phase, which causes a strong main phase peak of $\sim$100\%  above the background. 

According to previous studies, the multipolar magnetic fields may be a necessary condition for the production of an 
EUV late phase \citep[e.g.,][]{Hock12b,Liu13}. We propose a three-stage magnetic reconnection scenario to explain the flare 
evolution under such a magnetic topology. 

The main flare reconnection between N1 and  P2 is triggered by the tether-cutting mechanism \citep{Moore01},
as evidenced by the sigmoid-to-flux-rope eruption pattern. The appearance of the remote flare ribbon may suggest a secondary
null-point reconnection \citep{Masson09,Reid12}.  A magnetic null-point was indeed found by \citet{Jiang13}, but the 
outer spine field lines surrounding the null-point extend to the northwest rather than the eastern remote ribbon. Nevertheless, 
large-scale overlying field lines that connect P1 and N1 do exist. When the FR erupts, it strongly reconnects with and 
stretches these overlying field lines, causing the brightening of the remote ribbon and deep dimming at the far side 
of the remote ribbon. This FR-driven reconnection should belong to the breakout reconnection \citep{Antiochos99}.  

The eruption of the FR largely removes the overlying magnetic confinement. Therefore, the northern part of the filament starts to 
rise as a sequence of the torus instability \citep{Kliem06}. The rising filament also drives the breakout reconnection but in
a gentle manner, producing two sets of side-lobe post-flare loops (Figure 4(b)). Since the southern side-lobe loops located in the
main flare region are more closer to the reconnection site,  they are more prominent than the northeastern ones, responsible for the 
flare's intermediate phase.

As the filament-driven breakout reconnection turns the lower overlying field lines into side lobes, new reconnection occurs between 
the higher overlying field lines that are previously stretched. The process is similar to the main flare reconnection, but is much less 
energetic because of the weaker magnetic fields that are involved.  Hence the reconnection only produces moderately heated warm
EUV late phase loops (Figure 4(f)). These loops cool as they retract, sequentially brightening in cooler coronal channels. Meanwhile,
the whole  AR gradually recovers to its pre-flare state.  In a large-scale extent, the on-going reconnection can occur at different sites 
and with different rates, making the energy injection rather fragmented. The slow cooling rates for the late phase loops can 
be expected from their long length \citep[cf.][]{Cargill95}.

The three-stage magnetic reconnections show a casual relationship between each other, and the EUV late phase is mainly a 
production of the third stage reconnection. It is believed that some very early late phase loops (see Figure 4(b)) have already been
produced during the second stage breakout reconnection \citep{Hock12b}. The loops are not visible in cooler coronal channels just 
because they are still relatively hot at that moment.   
  
The diffuse morphology of the late phase loops seen in AIA 33.5 nm images is also reflected from the more smooth AIA 33.5 nm 
profile, as compared with the less smooth cooler emission profiles shown in Figure 4(i). \citet{Woods11} attribute this phenomenon
to the relatively slow cooling rate at warm coronal temperatures compared to that at cooler coronal temperatures. However, it seems 
not to be the case for the current event as revealed in Table 1. We present an alternative explanation in terms of the AIA TRFs. 
The AIA 33.5 nm TRF is quite flat from the peak with decreasing temperatures. It means that as a warm late phase loop cools its 
visibility in AIA 33.5 nm does not change much; therefore, there could be many overlapping loops visible at the same time. 
Nevertheless, the TRFs for the cooler coronal channels (AIA 21.1, 19.3, and 17.1 nm) have a sharp peak. In these channels, a small 
change of the loop temperature around the TRF peak will significantly change the loop intensity, resulting in the sharp loop 
morphology and the large fluctuation of the intensity profile.

\acknowledgements{We are grateful to the anonymous referee for his/her insightful comments.
This work is supported by NSFC (11103009, 10933003, 11203014, and 11078004), and by 973 project of China (2011CB811402).
\emph{SDO} is a mission of  NASA's Living With a Star (LWS) program.
}


\end{document}